**Strong Light-Matter Coupling Facilitated Charge Carrier Transport in Cavity Organic Solar Cells**

Yahui Tang,[1,2] Alexandra Stuart,[1] Timothy van der Laan,[2] and Girish Lakhwani [1*]

[1] ARC Centre of Excellence in Exciton Science, School of Chemistry, The University of Sydney, Australia

[2] CSIRO Manufacturing, Australia

*Corresponding author. Email: girish.lakhwani@sydney.edu.au

**Abstract**

Strong light-matter coupling has shown great potential for modifying the electro-optical properties of semiconducting materials in recent years. In the strong coupling regime, excitons and cavity photons form new states named exciton-polaritons, with their properties a hybrid of each constituent. Herein, we report strong coupling observed in solution-processed donor:acceptor bulk-heterojunction organic solar cells (OSCs) evidenced by the observed Rabi splitting of ~300 meV and the effects of strong coupling on OSC operations. Combining the transient photovoltage decay measurement and nanosecond transient absorption spectroscopy, our results reveal that the effective charge carrier lifetimes are longer in cavity devices, attributed to the reduced bimolecular recombination. It is also found that access to CT state(s) of higher energy is enabled in cavity devices. This study demonstrates that strong coupling can effectively modify the device- and photo-physics in OSCs and opens a new pathway for engineering more efficient OSCs.

**Introduction**

Strong light-matter coupling in organic microcavities is capable of modifying the electro-optical properties of molecular systems without changing their chemical structures.(*1, 2*) By placing molecules into an optical cavity in which the cavity mode is resonant with the optical transitions that form excitons in molecules, strong coupling between excitons and cavity photons can occur and produce new hybrid states named polaritons.(*3–6*) Polariton states feature properties from both their light and matter components, which makes them highly delocalized compared to the excitons (often Frenkel type) in organic semiconductors. (*7, 8*) It has been reported, for example, that polaritons in organic Fabry–Pérot cavities can propagate over hundreds of nanometres,(*9*) and surface Bloch wave polaritons can even propagate through ultrafast ballistic motion in the strong coupling regime.(*10*) In addition, strong coupling can lead to enhanced charge transport.(*11–13*) The boost in charge carrier mobilities of organic field-effect transistors under the strong coupling regime has been experimentally demonstrated for both n- and p-type organic semiconductors. (*12, 13*) These unique properties of polaritons have inspired a wide range of applications.(*14*)

One of the emerging applications, cavity organic solar cells (OSCs) in the strong coupling regime, has captured more attention in recent years. (*15–19*) The photoactive layer in organic solar cells typically consists of an electron-donating (donor) and an electron-accepting (acceptor) material. The blend solutions of two materials are spin-coated to form a bulk-heterojunction (BHJ) structure, in which the donor and acceptor are intermixed to achieve efficient photocurrent generation, with a schematic shown in Fig. 1(D). Photocurrent generation in BHJ organic solar cells includes four steps: (i) exciton formation following the photon absorption; (ii) exciton diffusion to the donor:acceptor interfaces; (iii) charge separation, including the formation of an intermediate charge transfer (CT) state and the subsequent dissociation into free charges; (iv) charge transport and collection. Strong coupling has the potential to improve OSC operations for multiple reasons. Firstly, with the polaritons exhibiting a longer diffusion length than excitons, they have a better chance of reaching the donor:acceptor interfaces, thus minimizing the loss at step (ii) through exciton recombination. (*15, 16*) Secondly, the new energy channeling pathway from the lower polariton state to the charge transfer state has been shown to be highly efficient in planar heterojunction donor:acceptor photodetectors and photovoltaics and can potentially improve step (iii).(*16*) Last but not least, the charge carrier transport may also be increased, since an increase in photoconductivity was observed for common OSC materials, such as P3HT and N2200, when they are they were coupled to a plasmonic cavity in a transistor configuration.(*12*, *13*) Studies on strong coupling in donor:acceptor heterojunctions has mostly focused on the ultrafast dynamics of the photophysical processes in the cavity,(*16*, *17*, *20*) while the slow dynamics of charges in cavity devices are yet to be investigated.(*21*)

Herein, we report the charge carrier dynamics in both working cavity devices and cavity samples (thin film sandwiched between planar silver cavities). The active materials used are PTB7-Th (donor) and $PC_{71}BM$ (acceptor), and the cavity is designed to couple the excitons generated on the donor material, which is a semi-crystalline conjugated polymer. There are multiple reasons why we chose to use these materials. The morphology of blend films of this system is well-understood and relatively uncomplicated, with diffraction measurements showing the molecular packing orientation of PTB7-Th remains almost unchanged compared to its neat form.(*22*) Also, the addition of common solvent additive DIO to the blend improves the intermixing of the donor and acceptor material, breaking the polymer phase into smaller domains.(*23*) Given that the absorption peaks of PTB7-Th and $PC_{71}BM$ are separated, this has a similar effect as diluting the polymer using a transparent matrix such as PMMA while maintaining a BHJ structure favourable for efficient charge separation and transport in organic solar cells. Taking all the above considerations, the chosen system is, therefore, an ideal platform for investigating the effect of strong light-matter coupling on photo-physics and device operations.

In this study, the evidence of strong coupling in the BHJ cavity samples and devices is presented, with distinct polariton peaks and dispersion behaviour observed in the angle-resolved reflectance measurements. The corresponding polariton peaks also appear in the external quantum efficiency spectra and the electroluminescence spectra of devices, suggesting that polariton states exist and, most importantly, are involved in the photocurrent generation in devices. The energy of the CT state(s) was found to be higher in cavity devices, as measured by electroluminescence and external quantum efficiency of devices. In addition, the charge carrier lifetimes of the reference and cavity devices were obtained from transient photovoltage (TPV) measurements and compared at the same charge carrier densities, with the charge carrier densities estimated using the transient photocurrent (TPC) measurement. Importantly, we found that the incorporation of polaritons leads to a longer effective charge carrier lifetime in cavity devices, which we attributed to reduced bimolecular recombination due to a smaller Langevin reduction coefficient, with the trend in TPV reproduced by transient drift-diffusion simulations. We verified that such longer charge carrier lifetimes are not due to resistance or capacitance effect in the electrical circuit, as the nanosecond transient absorption (TA) measurement on cavity film shows the same trend. This work demonstrates that strong coupling can be used to effectively manipulate the photo- and device physics of OSCs. Although it does not lead to an apparent efficiency increase in this case, because more than half of the light was blocked by the additional silver layer of 25 nm introduced to form the Fabry-Perot cavity, it is not an intrinsic limitation of the strong coupling phenomenon. Novel open cavity designs may be developed and used in the future to minimize this loss and improve overall device efficiencies.

## Results

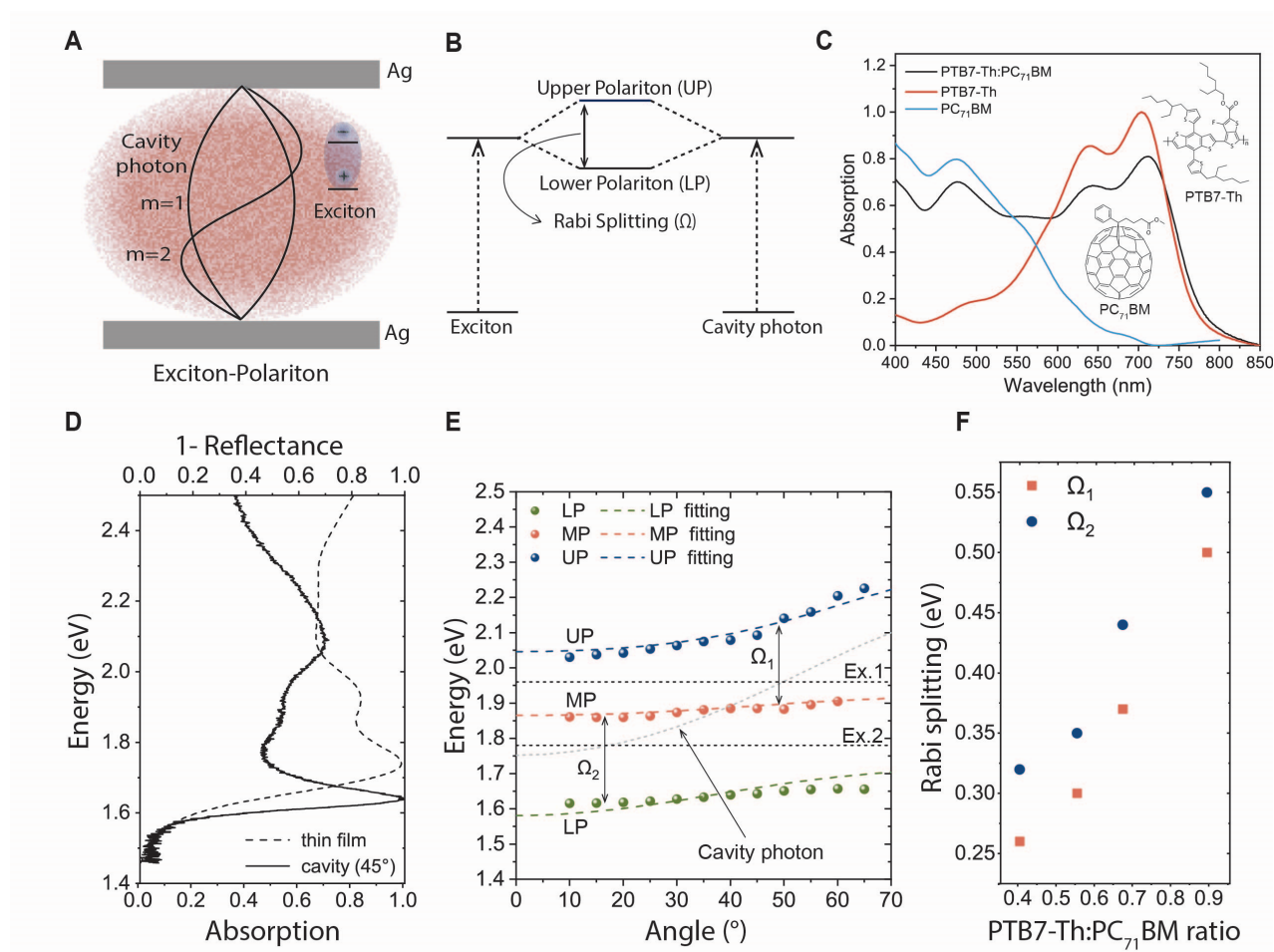

**Fig. 1. Strong coupling and bulk-heterojunction (BHJ) donor:acceptor organic solar cells.** Schematic of (**A**) exciton-polariton and (**B**) Rabi splitting. (**C**) Absorption of the neat PTB7-Th and $PC_{71}BM$ film, and blend film (PTB7-Th:$PC_{71}BM$ of 1:1.5 ratio). (**D**) The normalized absorption spectrum of the PTB7-Th:$PC_{71}BM$ blend film of 1:1.5 blend ratio, and the (1-Reflectance) spectrum of the corresponding cavity sample measured with a 45° incident angle. The cavity measured have a structure of Ag (25 nm)/PTB7-Th:$PC_{71}BM$/Ag (100 nm). The thickness of the PTB7-Th:$PC_{71}BM$ layer in the cavity is measured to be 170 ± 10 nm by a profilometer. (**E**) The dispersion curves of upper polariton (UP), middle polariton (MP), and lower polariton (LP) were measured by angle-dependent reflectance, along with the fitting to a coupled-oscillators model shown as dashed lines of the same colour corresponding to each polariton branch. (**F**) Ratio-dependence of Rabi splitting.

The donor and acceptor materials studied are PTB7-Th and $PC_{71}BM$, respectively, with their chemical structures and absorption spectra of spin-coated films presented in Fig.1(C). The donor material PTB7-Th is a semi-crystalline conjugated polymer with two distinct vibronic peaks at around 642 and 710 nm, implying the formation of HJ aggregates in the neat film.(*24*) The acceptor material $PC_{71}BM$ is a fullerene derivative, which mainly absorbs below 550 nm. The vibronic peaks of PTB7-Th remain the same in the PTB7-Th:$PC_{71}BM$ blend film of 1:1.5 ratio, which indicates the aggregation of PTB7-Th in the blend is similar to the neat film. It has also been reported that diffraction measurements show the molecular packing of PTB7-Th remains almost unchanged in the PTB7-Th:PC$_{71}$BM blend prepared with similar procedures, with PTB7-Th showing a face-on preferential orientation in the blend and neat films.(*22*)

The PTB7-Th: PC$_{71}$BM blend films are sandwiched between a thin Ag layer (25 nm) and a thick Ag layer (100 nm) to form a Fabry-Perot cavity. The thickness of the BHJ layer is designed for the cavity photon to couple with the PTB7-Th absorption peak. Compared to the absorption of bare blend films, the (1-Reflectance) spectrum of the cavity sample shows significant changes with distinct peak splitting around two vibronic features of PTB7-Th are observed as shown in Fig.1(D). Specifically, the resonant cavity shows three peaks corresponding to upper polariton (UP), middle polariton (MP), and lower polariton (LP) modes at 598 nm, 666 nm, and 758 nm, respectively. To confirm the system is in the strong light-matter coupling regime, we measured angle-resolved reflectance given that the polariton states are dispersive to the in-plane vector.(*25*) The full set of angle-dependent spectra are shown in Fig. S6, and the peak positions of each polariton state are extracted and plotted against the angle of incidence in Fig.1(E). As shown in Fig.1(E), three polariton branches demonstrate a clear anti-crossing behaviour, which can be fitted to a coupled oscillator model to extract the Rabi-splitting values. Good agreement between the data and fitting was achieved, with the Rabi-splitting values obtained as $\Omega_1 = 0.26$ eV and $\Omega_2 = 0.32$ eV, with $\Omega_1$ referring to the splitting between UP and MP, and $\Omega_2$ referring to the splitting between MP and LP, respectively. The value of $\Omega_2$ is larger than the linewidth of the corresponding excitonic peak (Ex.1) and cavity photon mode (see Fig. S2 and Table S2 for details). The value of $\Omega_1$ is close to the linewidth of the excitonic peak (Ex.1), but only LP is expected to interact with charge generation/process as the energy transfer from UP/MP is much faster than these processes.(*17*) The obtained PC$_{71}$BM is not coupled to the cavity in this case because the energy of cavity photons and the excitons do not match. This has been previously observed in a similar material, PCBM, which also did not show strong coupling at room temperature.(*17*) Experimental evidence for this is also shown in Fig. 2(F), where as the PTB7-Th content increases, the Rabi splitting increases accordingly.

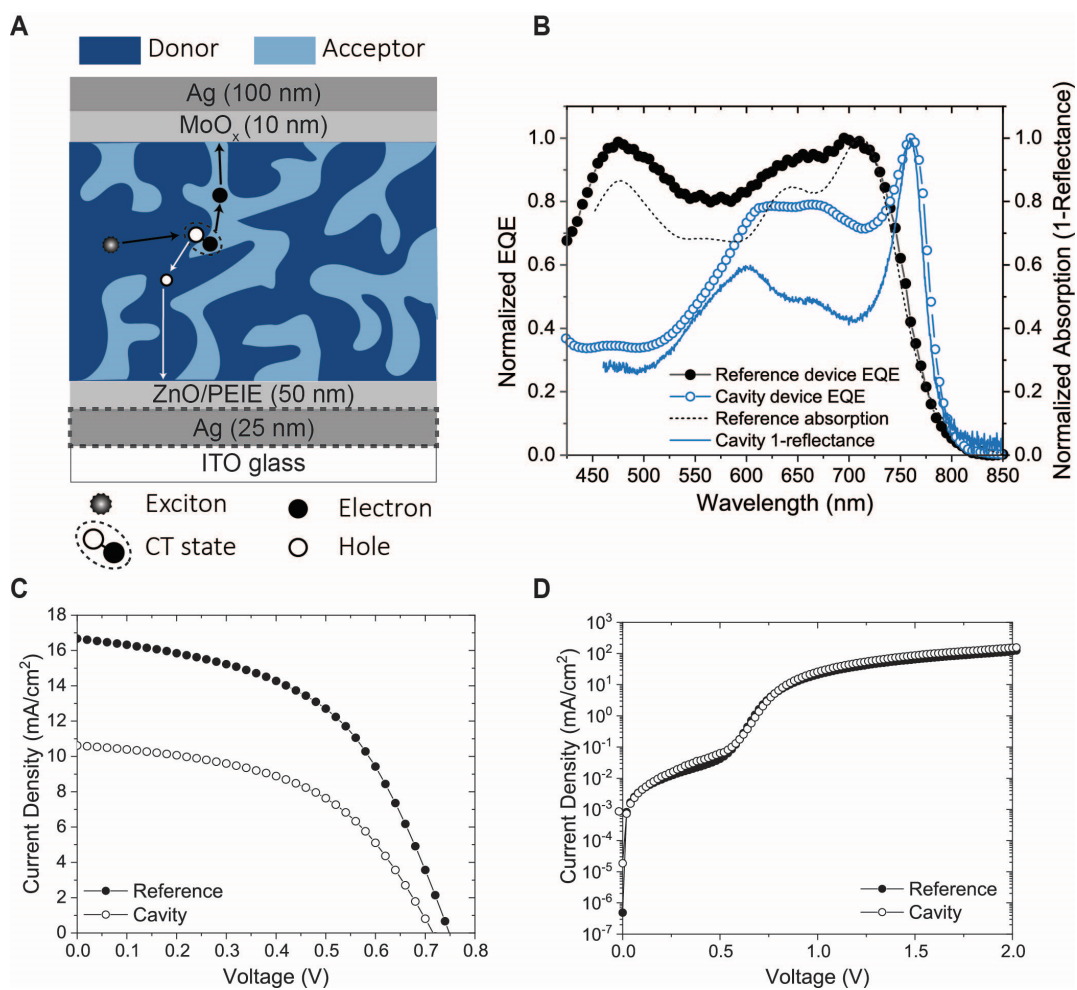

**Fig. 2. Photovoltaic performance of reference and cavity devices.** (**A**) The schematic of bulk heterojunction reference and cavity devices. The cavity device has an additional Ag layer of 25 nm compared to the reference. (**B**) The normalized external quantum efficiencies (EQE) of the cavity and reference organic solar cells, with three peaks corresponding to the upper polariton (UP), middle polariton (MP), and lower polariton (LP) states shown for the cavity device. (**C**) The light current density vs. voltage (JV) characteristics of the cavity and reference organic solar cell devices under 1 sun illumination. (**D**) The dark JV characteristics of the reference and cavity devices.

The photovoltaic performance of the reference and cavity device under AM 1.5G illumination is shown in Fig.2(A) and summarized in Table 1. The cavity device has an architecture of ITO/Ag/ZnO/PEIE/BHJ/MoO$_x$/Ag, with the BHJ layer having a thickness of $106 \pm 6$ nm in the resonant cavity device. The short-circuit current density ($J_{SC}$) of the cavity device is 10.6 mA/cm$^2$, which is lower than the reference device. This is expected because the upper Ag layer of 25 nm blocks more than half of the light. To quantify the optical loss caused by the thin Ag mirror, we performed transfer wave matrix simulations using the commercial software Setfos. This analysis provides a $J_{SC}$ of 7 mA/cm$^2$ if we only consider the optical effect introduced by the Ag mirror, that is, the optical loss from the upper Ag layer is around 60%. Details of the simulation are given in the supplementary materials. Comparing the simulated and measured values, the measured $J_{SC}$ of the cavity device is ~20% higher than that of the simulated $J_{SC}$ which only considers optical loss. This suggests that the strong coupling (not simulated) leads to enhanced photocurrent generation, which is not accounted for by the optical simulation. The open-circuit voltage ($V_{OC}$) of the cavity device is 0.04 V lower, but it is likely due to the decreased incident light intensity. The dark current density – voltage (JV) curves of the reference and cavity device are identical, as shown in Fig 2(D), which confirms the shunt resistance and series resistance are identical in both devices. Similar to the reflectance spectra, three peaks corresponding to polariton branches are also observed in the spectral response of the cavity device measured as the external quantum efficiency (EQE) spectrum. As shown in Fig. 2(B), three peaks corresponding to the UP, MP, and LP appear at 598 nm, 666 nm, and 758 nm, respectively. Unlike in the reference device, the current generation of the cavity device is mostly from the polariton peaks rather than the $PC_{71}BM$ absorption. The unnormalized EQE in Fig.S4 shows that the EQE of the cavity device is larger than the reference device in the region ~760–790 nm, which corresponds to the LP peak, while it is lower in other wavelength regions due to the optical loss caused by Ag layer of 25 nm. In contrast to the polariton region, the EQE in the region corresponding to $PC_{71}BM$ absorption is reduced to around 30% of the reference device (see Fig.S4 in supplementary materials).

**Table 1**. Photovoltaic parameters of the cavity and reference devices. The averaged values and standard deviations are from 5 devices.

| Device | $J_{SC}$ (mA/cm²) | *Simulated $J_{SC}$ (mA/cm²) | $V_{OC}$ (V) | Fill Factor | Efficiency (%) |
|---|---|---|---|---|---|
| Reference | 18 ± 1 | 17 | 0.77 ± 0.01 | 0.56 ± 0.04 | 8.0 ± 0.5 |
| Cavity | 13 ± 2 | 7 | 0.73 ± 0.01 | 0.49 ± 0.04 | 4.6 ± 0.9 |

*Transfer wave matrix simulations were conducted in commercial software Setfos to estimate the optical loss caused by the additional Ag (25nm) layer in cavity devices. The polariton effect is not simulated.

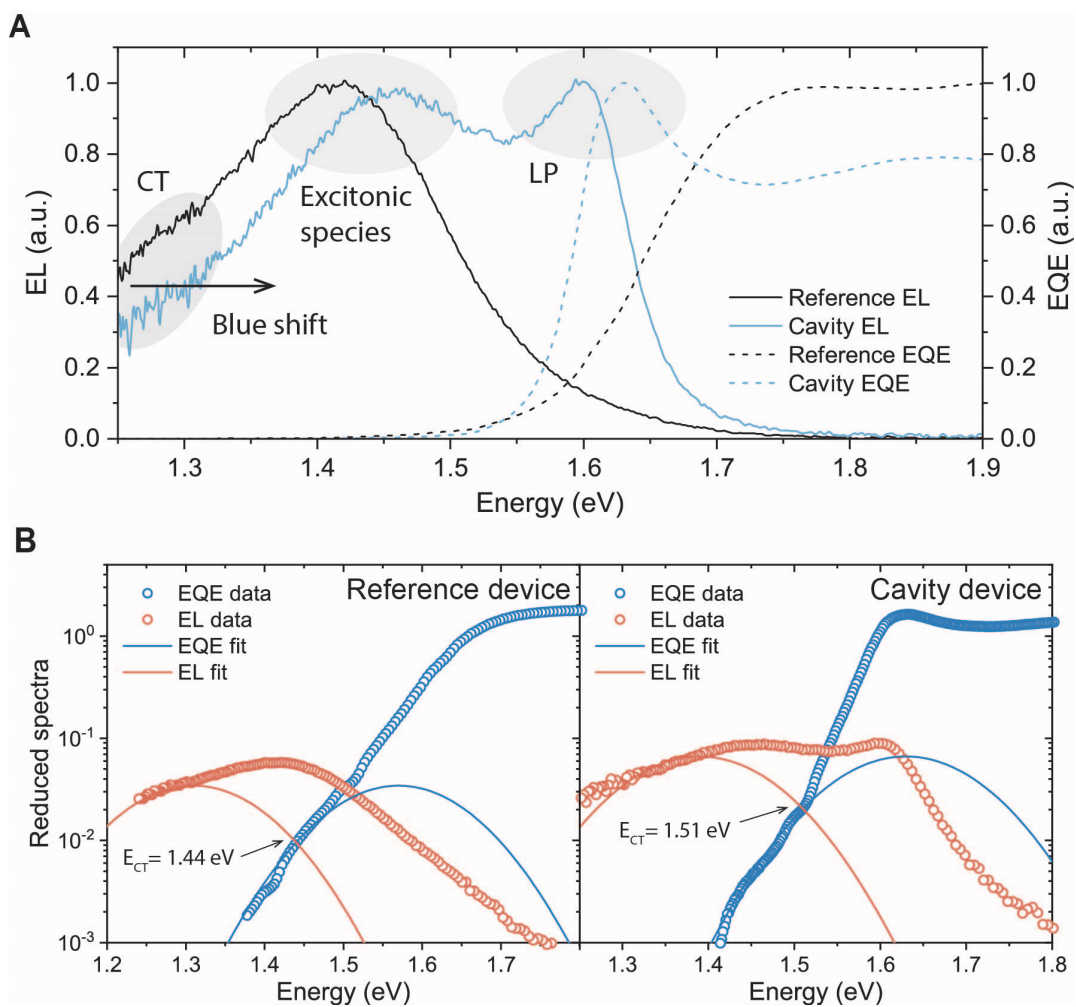

**Fig. 3. Electroluminescence and $E_{CT}$ of reference and cavity devices.** (**A**) External quantum efficiency (EQE) and Electroluminescence (EL) of devices under linear scale. (**B**) Fitting of EQE and EL for extracting the energy of CT state ($E_{CT}$), with all the fitting parameters summarized in Table S3.

The electroluminescence (EL) spectrum of the cavity device also differs significantly from that of the reference device, as shown in Fig.3(A). The lower polariton peak appears in the electroluminescence (EL) spectrum of the cavity device, symmetric to the LP peak in the EQE spectrum with respect to the intersection of the EQE and EL. The peak at around 1.45 eV is more likely emission from excitonic species, similar to that of the peak (at ~ 1.4 eV) observed for the reference device but showing a blue shift of ~0.05 eV. The regime below the exciton peak in the EL spectrum of the reference device can be attributed to the emission of charge transfer (CT) state, as the same system has been studied and reported by multiple references. (*26, 27*) Such a regime is also observed in the cavity device, which we, therefore, attribute to the emission of the CT state. Compared to the reference device, the cavity device shows a blue shift of EL in the CT regime (the feature at ~1.3 eV), suggesting there is a change in the energy of the CT state ($E_{CT}$). To estimate the change of $E_{CT}$, the region assigned to CT state in EQE and EL spectra are simultaneously fit into two Gaussian line shapes,(*28*) obtaining values of $E_{CT}$ for both the reference and cavity devices. The fitting for both the reference and cavity device is shown in Fig.3(B), with the $E_{CT}$ used as a shared fitting parameter during the fitting of both EL and EQE spectra and is determined as the cross point of the fitted Gaussian lines. More details about the fitting are given in Fig.S8 and Table S4. The obtained $E_{CT}$ is 1.44 eV for the reference device, similar to the values reported by the literature.(*26*, *27*) The obtained $E_{CT}$ for the cavity device, in contrast, is 1.51 eV, which is 0.07 eV (~ 3 $k_BT$) higher than that of the reference, suggesting that the access to charge transfer state(s) of higher energies are enabled in the cavity device.

In addition, effective charge carrier lifetimes are found to be longer in cavity devices. The charge carrier lifetimes are determined by transient photovoltage (TPV) measurement and calibrated to the charge carrier densities measured using transient photocurrent (TPC) at different intensities.(*29*, *30*) Fig. 4(A) and (B) show the intensity-dependent decay of photovoltage of the reference and cavity devices, respectively. The decay lifetime of both devices is dependent on the intensity of incident light, which is expected because the dominant recombination mechanism of PTB7-Th:PC$_{71}$BM devices at the open-circuit condition is bimolecular recombination, which is dependent on the product of charge carrier densities. Since the upper 25 nm Ag mirror in the cavity device blocks more than half of the incident light, it is thus important to compare the charge carrier lifetime at the same charge carrier densities rather than just relying on the absolute intensity of incident light. TPC was thus measured at different intensities, with the intensity-dependent charge carrier densities obtained by integrating the decaying tail after the light is turned off. The intensity-dependent TPC curves and the intensity-dependent extracted charge carrier intensities are given in Fig.S9, and the obtained charge carrier density is shown in Fig. 4(C). The obtained charge carrier lifetimes from TPV are plotted against the charge carrier densities from TPC of the same light intensities to present an effective comparison between the charge carrier lifetimes of reference and the cavity device in Fig. 4(D). Significantly, the charge carrier lifetime is longer in the cavity device when the charge carrier densities are equal, which suggests that the recombination rate of charges is lower in the cavity device, which features polariton states. The bimolecular recombination mechanism of PTB7-Th:PC$_{71}$BM is reduced Langevin recombination, with the bimolecular recombination rate described by:(*31*)

$$R = \frac{\gamma(\mu_n+\mu_p)q}{2\varepsilon}(np - n_i^2) \qquad \text{Equation 1}$$

in which the bimolecular recombination rate (R) is reduced by multiplying a Langevin reduction coefficient (γ) compared to an ideal Langevin recombination rate. Other parameters in the equation include electron/hole carrier mobilities ($\mu_{n/p}$), electron/hole densities ($n/p$), and intrinsic carrier densities ($n_i$). Typical values of γ are < 0.2, with a smaller γ leading to a lower recombination rate, and thus a longer charge carrier lifetime in TPV measurement. We demonstrate the effect of varying γ on the TPV measurement in Fig.4(E) by transient drift-diffusion simulation, in which we show that a larger γ indeed leads to a slower decay in TPV measurement. Parameters used for simulation are presented in Table.S3, and the physical meaning of this observation will be discussed later.

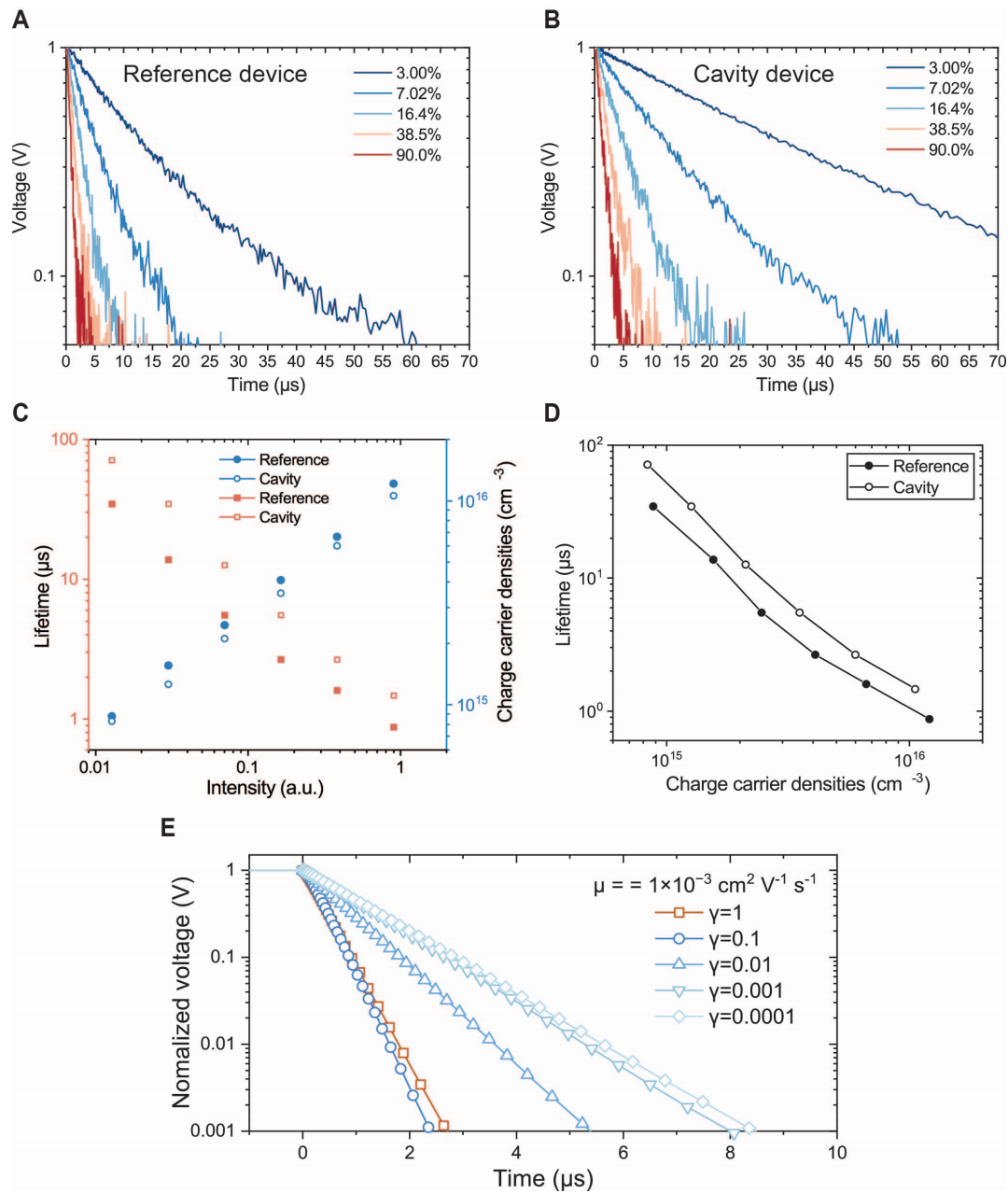

**Fig. 4. Longer charge carrier lifetime in the cavity device.** Intensity-dependent transient photovoltage decay of (**A**) reference device and (**B**) cavity device. (**C**) Intensity-dependent

charge carrier lifetime and densities. The charge carrier lifetime was estimated by integrating the tails of transient photocurrent decay. (**D**) The lifetime comparison of reference and cavity devices at different charge carrier densities. (**E**) Simulation showing the effect of reduced Langevin reduction coefficient ($\gamma$) in TPV measurements. As $\gamma$ decreases, the decay becomes slower in TPV measurements, showing longer effective charge carrier lifetimes.

We verified the effect of strong coupling using nanosecond transient absorption (TA). Fig. 5(A) and (B) show TA spectra of a reference sample (bare blend film) and cavity sample, exciting at 532 nm. The cavity sample used for TA had 25 nm Ag on both sides, in order to collect TA in both transmittance ($\Delta A$) and reflectance ($\Delta R$) modes. Further, TA data of the cavity sample measured in reflectance mode at 45° and transmittance mode at 0° is shown in the Supplementary Material, along with TA data of neat PTB7-Th. The reference sample shows a PTB7-Th ground-state bleach (GSB) around 730 nm, and an ESA around 1200 nm that has previously been attributed to the absorption of CT states/charge separated (CS) states,(*32*) or hole polarons.(*33*) CS states and hole polarons are similar in this context and cannot be distinguished in TA, we therefore consistently address these states as charges in the following discussion. There is additionally a small GSB due to $PC_{71}BM$ at 470 nm. At the excitation fluences used for TA in this study, neat PTB7-Th films, reference sample (blend film), and cavity sample (blend cavity) all show significant fluence-dependent spectra and decays, indicating the presence of bimolecular processes. For blend and neat PTB7-Th films, higher excitation fluences result in a significantly faster decay of the CT state/charges ESA, as well as the formation of an ESA at 600 nm due to an unknown species (SM Figures 2 and 5 in TA part). The exact mechanism of the formation of this species is unclear. However, we observe that in blend films, less of this species is formed than in neat PTB7-Th, and the CT state/charges ESA lives longer at equivalent fluences, indicating that the bimolecular process(es) is suppressed upon the addition of $PC_{71}BM$. Further detail on bimolecular processes in non-cavity films can be found in the supplementary materials.

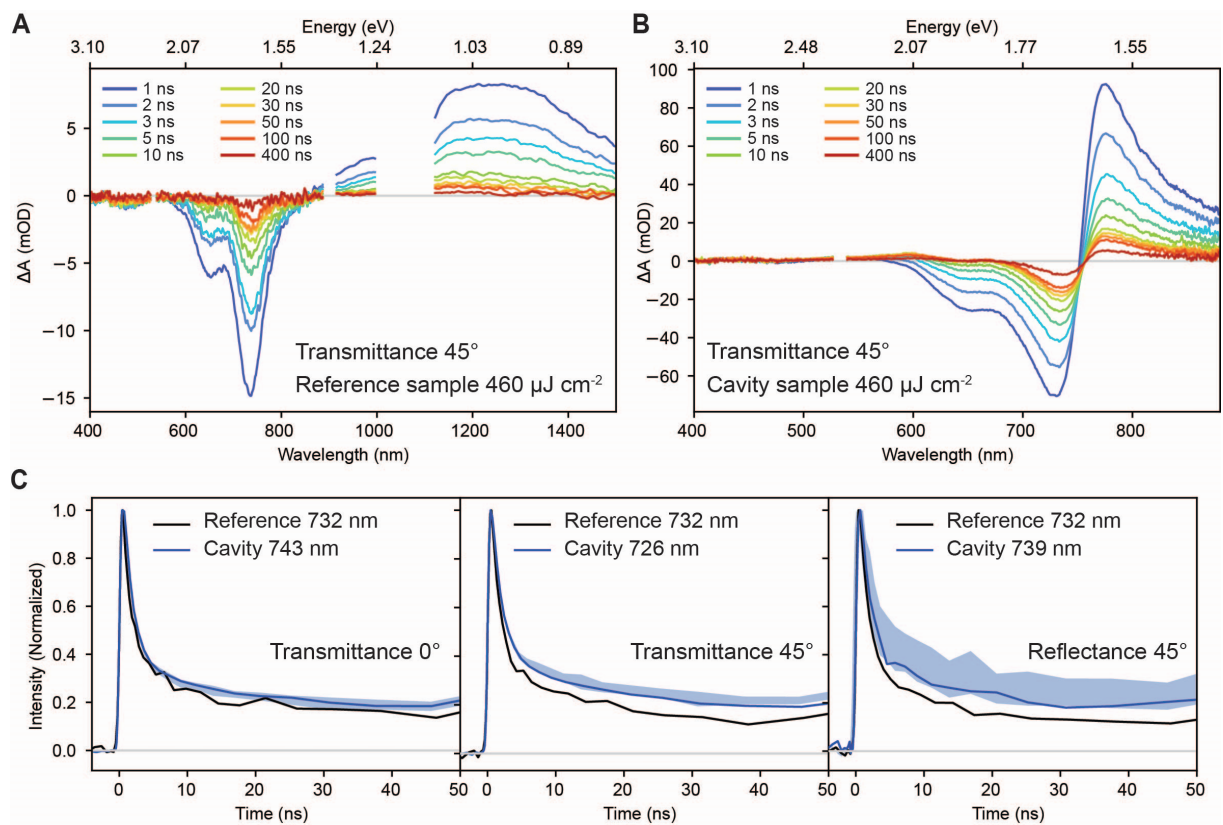

**Fig. 5. Polariton effect in TA of resonant cavities.** Nanosecond TA spectra of PTB7-Th: PC$_{71}$BM of 1:1.5 blend ratio (**A**) reference sample and (**B**) cavity sample, exciting at 532 nm and 460 μJ cm$^{-2}$. The cavity sample has a structure of Ag(25m)/BHJ/Ag(25nm). The full spectra shown were measured at 45° from the normal to the sample. The kinetic traces in (**C**) show the decay of the ground state bleach at 0° and 45° in transmittance mode, and at 45° in reflectance mode. The cavity sample was not measured in the NIR due to poor probe transmittance in that region.

The TA spectra of the PTB7-Th:PC$_{71}$BM cavity sample differ from the reference sample. Most notably, the PTB7-Th GSB is obscured by a derivative-like feature around 750–800 nm. Similar features have been observed in numerous cavity transient absorption studies, and in part correspond to a phenomenon referred to as polariton contraction.(*34–36*) The strength of the light-matter coupling, and thus the energies of upper and lower polaritons, are proportional to the number of chromophores in the system, in this case PTB7-Th chromophores in the ground state. After the cavity is excited, there are fewer ground-state chromophores, causing a reduction in the Rabi splitting and a "contraction" of polariton energies.(*36*) This contraction causes changes in the ground-state absorption or reflectance spectra with excitation, which results in derivative-like signals in the TA spectrum. Changes in the ground-state absorbance and reflectance also arise from excitation-induced changes in cavity length or refractive indices.(*36*) The positive signal around 800 nm observed in the cavity TA, therefore, does not necessarily correspond to an excited-state absorption. The dynamics and decay of this feature should be representative of the total excited state or charge population, in the same way that the dynamics of the bare film GSB represent all excited states and charges.

Given that there is clearly excitation fluence dependence in both reference and cavity samples, it is of interest to determine whether the lifetimes of the CT states/charges are changed in the cavity. Fig.5(C) shows the decay of the PTB7-Th GSB in the reference sample and the derivative feature in the cavity sample, excited at 460 µJ cm$^{-2}$. This excitation fluence was chosen as it is high enough to show some bimolecular recombination, but low enough that the decay of the GSB still primarily reflects the decay of hole polarons, rather than the 600 nm absorbing product. The decay was measured at this excitation fluence at 0° and 45°. At 0°, the decay of the bare and cavity film is essentially the same within error. However, at 45°, when the cavity is resonant with the PTB7-Th absorption, the cavity decay is slower in both transmittance and reflectance modes. The cavity was measured at higher incident excitation fluence to ensure that the internal fluence experienced by the active layer was comparable to the reference sample. There is a level of uncertainty in the transmittance of the Ag layer, so the cavity was measured at the upper and lower bounds of the incident excitation fluence corresponding to an internal fluence of 460 µJ cm$^{-2}$, giving the shaded blue areas in Fig.5(C). The actual internal fluence of the cavity is likely larger than this, as the bottom Ag layer will reflect some of the excitation pulses, providing a second chance for the active layer to absorb it. Since this would only increase the cavity decay further, we can conclude that at the same or even higher fluences, the cavity decay is slower. This result is consistent with the longer lifetimes observed for cavities in TPV measurements, although the absolute magnitudes of the fluence and lifetimes are different. The cavity sample used in TA has a different configuration, and TA is measured under different conditions (i.e. in the absence of an external field and charge transporting layers), so we do not expect the magnitudes of these values to be comparable.

# Discussion

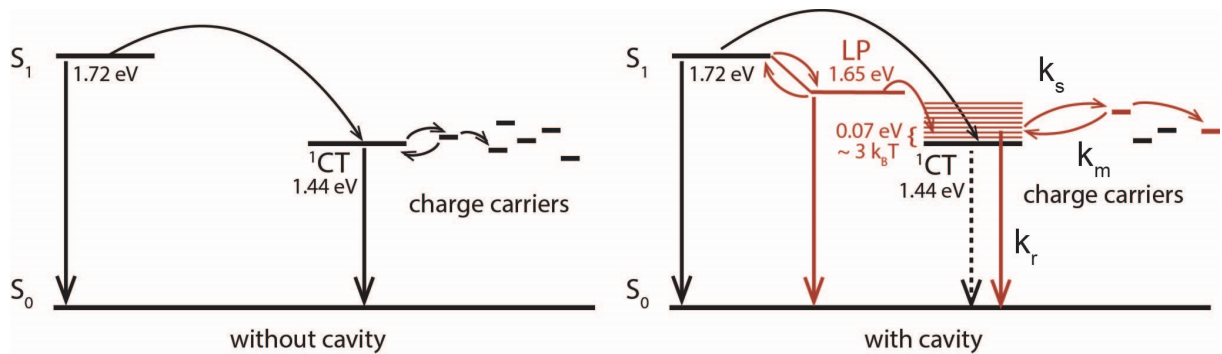

**Fig. 6. Polariton effects.** Compared to the reference device, the cavity device mainly shows three changes: (i) The polariton states are involved in the photocurrent generation, opening a new charge generation pathway. (ii) Although the lower polariton has lower energy than the excitons (mainly singlet state in this donor:acceptor system), the CT state shows a 70 meV blue shift. (iii) The longer charge carrier lifetime measured in TPV and TA, which is likely a result of lower CT relaxation rate $k_r$ or faster CT dissociation rate $k_s$.

The previous studies on cavity donor:acceptor OSCs mainly focused on ultrafast dynamics and aimed to directly probe the formation of charges from the new channel opened by the polariton states. Specifically, DelPo et al. identified the additional charge generation pathway from polaritons,(*17*) and Wang et al. reported that the rate of energy transfer is increased two- to threefold in the strong coupling regime.(*16*) We instead studied the slow photophysical processes by measuring the nanosecond transient absorption of cavity samples as well as the transient photovoltage (TPV) decay of functioning devices. Such a slow process can be interpreted as the dynamics of recombination/transport of charge carriers. Significantly, we found that the charge carrier lifetime is longer in the cavity, with nanosecond TA and the TPV showing a consistent trend.

The physical implications of our results are discussed below. Since the dominant recombination in the studied system (namely PTB7-Th:PC$_{71}$BM) is reduced Langevin recombination, the decay of charges is limited by two processes: (i) the encounter of charges/formation of the CT; (ii) the subsequent relaxation of the CT state.(*29, 30, 37*) The formation rate of CT is dependent on the charge carrier densities and can be described as Langevin type recombination, while the possibilities of CT relaxation after formation is determined by the competition between the re-dissociation and the CT relaxation,(*38*) as illustrated in Fig.6. The Langevin reduction coefficient $\gamma$ is usually added to account for such an effect in device modelling as shown by Equation 1, and it is argued that $\gamma \sim \frac{k_r}{k_r+k_s}$ by Burke et al.(*38*) Therefore, the longer charge carrier lifetime observed in TPV and TA is likely a result of a fast CT dissociation rate and/or a slower CT relaxation rate, which leads to longer effective charge carrier lifetimes in TPV and TA measurements. This might be related to the blue shift of the CT state measured by the EQE in the sub-bandgap region (CT absorption) and the electroluminescence in the infrared region (CT emission). Although the exact origin of the blue shift is hard to determine here, such a change observed in the cavity is not strange. A similar phenomenon was observed experimentally before in all-optical pump-push-probe spectroscopy measurements on polymer:fullerene blends. (*39*) In these measurements, an additional push (an infrared optical excitation) was applied to recover the excited, delocalized band states which are higher in energy than the lowest-lying CT state and can facilitate long-range charge separation. (*39*) As theoretical calculations predict, the cavity-induced polaron decoupling in the strong coupling regime can modify the photophysical response of molecular aggregates that involve nuclear rearrangements in the excited states. (*40*) It is therefore likely that the polaron decoupling in cavities can realize the same effect in working devices as applying a push pulse in spectroscopic measurements of films.

Our results show that strong coupling can lead to a lower Langevin reduction coefficient in PTB7-Th:PC$_{71}$BM cavity OSC devices, which is a potential pathway for improving photovoltaic performance. To evaluate the benefits of a smaller Langevin reduction coefficient, we performed steady-state drift-diffusion simulations and found that the improvement can indeed be significant, especially for the devices that have lower mobilities. As shown in Fig. S24, a smaller $\gamma$ can lead to a substantial increase in both $J_{SC}$ and $V_{OC}$, especially for low mobilities systems. For systems with mobilities of $10^{-4}$ cm$^2$ V$^{-1}$, a 1.5 times increase in the power conversion efficiency (PCE) is predicted in the simulations decreasing $\gamma$ from 0.1 to 0.001, as shown in Fig. S24(D). Indeed, it has been observed recently that some non-fullerene acceptor-based systems have a smaller Langevin reduction coefficient compared to the fullerene-based systems, which leads to a higher fill factor and lower recombination loss.(*41*) While material design and synthesis can be a potential way of realizing a smaller Langevin reduction coefficient, our results demonstrate that strong coupling can be an alternative approach for this purpose.

To conclude, strong light-matter coupling is demonstrated in BHJ donor:acceptor OSCs and is shown to modify the photo- and device physics. In particular, we found that the overall photocurrent generation is enhanced in these devices if the optical loss introduced by the thin silver layer is accounted for. More importantly, we show that the CT state shows a blue shift of around 70 meV, along with a longer charge carrier lifetime in the cavity device. The longer charge carrier lifetime is attributed to a smaller Langevin reduction coefficient, which has the potential to lead to significant efficiency improvement for OSCs of low mobilities according to drift-diffusion simulations. From the perspective of engineering, replacing the planar metal mirror with other open cavity designs could reduce the optical loss and lead to a pathway toward an apparent power conversion efficiency increase in cavity solar cell devices.

**Materials and Methods**

**Materials**: PTB7-Th (#YY22138CH) and $PC_{71}BM$ (#MY11007PI) were purchased from 1-Material Inc.

**Device fabrication**: ITO glasses were cleaned subsequentially using acetone and isopropanol in an ultrasonic bath for 15 minutes each, followed by 15 min UV-ozone treatment. The inverted devices have ITO/Zno/PEIE/BHJ/$MoO_x$/Ag structures. The ZnO was spin-coated on the cleaned ITO substrates using half-diluted ZnO nanoparticle ink purchased from Infinity PV (5.6% w/v in IPA), producing ZnO layers of around 40 nm. The nanoparticle ink was filtered using 0.45 $\mu m$ PTFE filters before use. The ZnO films are annealed at 120 °C for 2 min after spin-coating. After the ZnO films preparation, PEIE(Polyethylenimine) layer was spin-coated on top using 0.1 wt% PEIE in IPA. The PTB7-Th:$PC_{71}BM$ layer was spin-coated on top of the ITO/ZnO/PEIE substrate from the solution of 1:1.5 ratio dissolved in dichlorobenzene (DCB) and 3% solvent additive 1,8-diiodooctane (DIO). After the active layer is deposited, the $MoO_x$ layer (10 nm) and Ag layer (100 nm) are thermally evaporated in a vacuum evaporator.

**Device characterization**: The Current density-voltage (JV) curves fabricated devices are first measured using a source-measure unit (Keithley 2450) under AM1.5G simulated sunlight produced by a solar simulator (Oriel LCS-100TM Small Area Sol1) with the intensity calibrated by a silicon reference cell. The device parameters were then extracted from the measured JV curves. The external quantum efficiency (EQE) was then measured using an integrated system IQE 200B from Oriel Instruments.

**Steady-state optical measurement**: The steady-state absorption was measured by the Cary-60 spectrophotometer, and the steady-state emission was measured by the Cary Eclipse fluorescence spectrometer. The angle-resolved reflectance was measured by varying the angle of incident white light (NKT, Compact) using a goniometer and then collected with an integrating sphere, which is fibre-coupled to an Ocean Optics Flame spectrometer.

**Transient absorption**: Transient absorption (TA) measurements were performed on a commercial nanosecond-timescale TA spectrometer (Ultrafast systems, EOS). The excitation pulse was a 532 nm pulse at 1 kHz repetition rate (Innolas) with a spot size of 0.4 mm at the sample and with polarization rotated to magic angle relative to the probe pulse. The probe pulse was generated from a photonic crystal fiber-based supercontinuum laser (Ultrafast systems, EOS). The probe pulse was split into a signal and reference beam and focused onto the sample with a 0.1 mm FWHM spot size. The probe was set at either 0° or 45° from the normal to the plane of the sample. The excitation pulse was less than 5° away from the probe pulse.

**Transient photocurrent (TPC) and transient photovoltage (TPV)**: TPC and TPV were all measured using the all-in-one characterization platform, Paios (Fluxim AG, Switzerland). The white LED that comes with the Paios setup was used for both measurements. Square light pulses of 500 μs were used during TPC measurement. In TPV measurement, a resistor of 1MΩ was used so that the device is under the open-circuit condition. During the measurement, a small light pulse (10%) was applied to offset light intensities, and the offset intensity-dependent voltage decay was recorded.

**Acknowledgments**
The authors thank Dr. Trevor Smith, Dr. James Hutchinson, and Dr. Rohan Hudson from The University of Melbourne, and Dr. Anthony Chesman from CSIRO for helpful discussions. Y.T., A.S. and G.L thank the funding support by the Australian Research Council Centre of Excellence in Exciton Science (funding grant number CE170100026)

**Funding:** Australian Research Council Centre of Excellence in Exciton Science CE170100026